\documentclass[%
 reprint,
 amsmath,amssymb,
 aps,
]{revtex4-2}

\usepackage{graphicx}
\usepackage{dcolumn}
\usepackage{bm}

\newcommand{\be}[0]{\begin{equation}}
\newcommand{\ee}[0]{\end{equation}}
\newcommand{\bea}[0]{\begin{eqnarray}}
\newcommand{\eea}[0]{\end{eqnarray}}

\newcommand{\ud}[0]{{\rm d}}
\newcommand{\ii}[0]{{\rm i}}

\newcommand{\Tr}{\text{Tr}}

\begin{document}

\preprint{APS/123-QED}

\title{Coherence effects on estimating general sub-Rayleigh object distribution moments}

\author{Kevin Liang}
 \email{kevinqcliang@gmail.com}
\author{S. A. Wadood}%
\author{A. N. Vamivakas}
    \email{nick.vamivakas@rochester.edu}
    \altaffiliation[Also at ]{Department of Physics and Astronomy, University of Rochester, Rochester, NY 14627, USA}
    \altaffiliation[Also at ]{Materials Science, University of Rochester, Rochester, NY 14627, USA}
\affiliation{The Institute of Optics, University of Rochester, Rochester NY 14627, USA\\
 Center for Coherence and Quantum Optics, University of
Rochester, Rochester, NY 14627, USA  
}%

\date{\today}

\begin{abstract}
Quantum Fisher information and signal-to-noise ratio bounds are derived for the estimation of moments of general partially coherent objects. Under an asymptotic analysis in the sub-Rayleigh regime, these bounds are shown to be less restrictive than those of the incoherent case. This leaves open the possibility of more accurately estimating object distribution moments when the object is partially coherent. 
\end{abstract}

\maketitle


\section{Introduction}

For realistic imaging systems, the ability to accurately and fully describe a general object scene deteriorates as the scene's characteristic size approaches the sub-Rayleigh regime. Within this regime, whose domain is determined by the width of the imaging system's point spread function (PSF), the prominent spatial frequency components of the object fail to transfer faithfully to the image plane and, as a result, the resulting image is irrecoverably blurred in a passive imaging system. The maximum spatial frequency components that successfully pass through to the image plane define, roughly, the resolution of the imaging system. Therefore, it is desirable, and indeed an ongoing area of research, to find ways to obtain \textit{superresolution}, which are methods of imaging that circumvent the aforementioned limitations in the sub-Rayleigh regime. These techniques include stochastic optical reconstruction microscopy (STORM) \cite{hell2007review}, stimulated emission depletion microscopy (STED) \cite{hell2007review}, and confocal microscopy with superoscillations \cite{GburReview,Smith2016, Aharonov1988,Kosmeier2011}. 

However, there has recently been an increase in interest regarding superresolution techniques based on modal decomposition \cite{Tsang2016, Tsang2018, Bearne2021}. The foundations of these ideas were first popularized by Tsang in the context of estimating the separation between two incoherent point sources \cite{Tsang2016}. The theoretical backbone of the analysis was the quantum Fisher information (FI), which provides a measurement-free information lowerbound on parameter estimation \cite{Helmstrom}. The remarkable result that the quantum FI for incoherent two-point separation remains positive even as the separation vanishes [along with a proposal of an measurement scheme known as spatial mode demultiplexing (SPADE)] paved the road for implementations and generalizations of this FI-based superresolution. For the two-point separation problem, these include experimentally demonstrating the proposed sub-Rayleigh precision \cite{OptExpress_Parity_Sorting2016,sanchezsoto_2016_optica,superresolution_with_heterodyne,steinberg2017beating_rayleighscurse}, extending the theory to include axial separation, and analyzing the effects of including additional parameters, such as centroid and relative intensities \cite{yiyu2019axial_superresolution_optica, Rehacek2017}. 

The extension of the quantum FI formalism to include the coherence between two point sources has sparked significant debate. This effort included the analysis of point sources that are partially coherent with a real coherence parameter \cite{Larson2018, Wadood2021}, fully coherent \cite{Hradil2019}, and arbitrarily partially coherent \cite{Liang2021}. Although various results have been challenged, primarily due to disagreements in how the quantum FI should be defined, it is clear that the behavior of the quantum FI changes appreciably when the two point sources are no longer incoherent.

Although the case of two point sources gives perhaps the clearest definition of resolution and allows the theory to be mathematically tractable, realistic objects are often more complicated to characterize. To this end, work has been done for finding the limits of fully characterizing general sub-Rayleigh incoherent object distributions \cite{Zhou2019, Tsang2019}. Furthermore, despite a lack of a complete theoretical understanding for their benefits in resolution, simulation results of techniques that utilize modal decomposition in conjunction with traditional imaging have demonstrated an improvement in characterizing complex incoherent object scenes \cite{Bearne2021}.

In this work, the quantum FI bounds for characterizing a general object distribution with arbitrary coherence properties are derived using moments. We find that a suitable starting point is the coherent mode decomposition (CMD) and parametrizing the estimation problem in terms of the moments of each coherent mode. These results may then be reinterpreted for the moments of the object mutual coherence and the object intensity distribution through reparametrization techniques. We find that the derived information bounds leave room for improvement with partially coherent sources when compared to the incoherent case.

\section{Theory} \label{theory}

\begin{figure}
    \centering
    \includegraphics[scale=0.7]{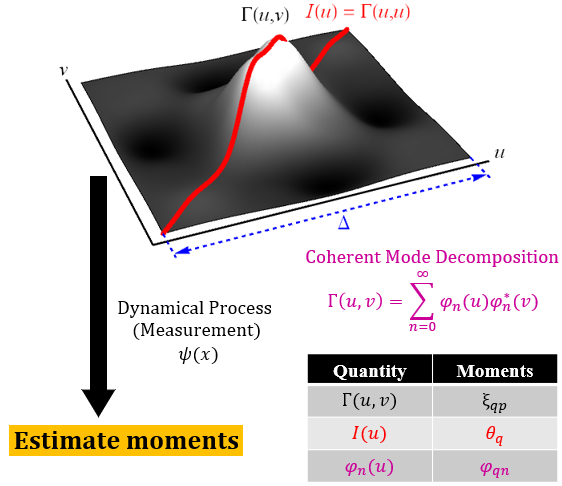}
    \caption{The object mutual coherence function, $\Gamma$, with characteristic size $\Delta$, is shown over the object-plane spatial coordinates $(u,v)$. The object intensity distribution (red), $I(u) = \Gamma(u,u)$, is a particular cross-section of the mutual coherence function. An alternative representation of $\Gamma$ is given by the coherent mode decomposition, with modified coherent modes $\varphi_n$. $\Gamma$ passes through an (non-unitary) dynamical process with PSF $\psi(x)$, and a measurement is performed at the image plane, from which $(\xi_{qp},\theta_q,\varphi_{qn})$, the moments of $(\Gamma, I, \varphi_n)$, respectively, are estimated.}
    \label{schematic2}
\end{figure}

The density matrix that represents the image of a general partially coherent object distribution passing through an imaging system is given by
\begin{align}
    \hat{\rho}_\alpha = \mathcal{N}_\alpha^{-1} \iint_{-\infty}^\infty \Gamma(u,v) e^{-\ii \hat{k}u} |\psi\rangle \langle \psi| e^{\ii \hat{k} v} \, \ud u \, \ud v, \label{densMat}
\end{align}
where $\Gamma(u,v)$ is the mutual coherence function of the object distribution as a function of two object-plane coordinates. Note that $\Gamma$ is Hermitian; that is, $\Gamma(u,v) = \Gamma^*(v,u)$. Furthermore, 
\begin{align}
    |\psi \rangle \triangleq \int_{-\infty}^\infty \psi(x) |x\rangle \, \ud x = \int_{-\infty}^\infty \Psi(k)|k\rangle \, \ud k,
\end{align}
where $\psi$ and $\Psi$ are the imaging system's (coherent) PSF and transfer function, respectively. They are respectively functions of the conjugate variables $x$ and $k$, which are the image-plane position and spatial-frequency coordinates. The collections of kets $\{|x\rangle\}$ and $\{|k\rangle\}$ each form an orthonormal basis and the overlap $\langle x | k \rangle = e^{\ii k x}$ indicate that $\psi$ and $\Psi$ are a Fourier-transform pair. A full derivation of Eq.~(\ref{densMat}) is provided in Appendix~\ref{appA}. Finally $\hat{k}$ is the one-photon spatial-frequency, or momentum, operator; both $\hat{k}$ and the object-plane posiiton coordinates $(u,v)$ are normalized with respect to the system's magnification and the bandwidth of $\Psi(k)$. This allows for them to be interpreted as unitless quantities. A schematic of some of the aforementioned quantities are shown in Fig.~\ref{schematic2}

Note that $\hat{\rho}_\alpha$ is indexed by $\alpha \in \{0,1\}$, which reflects the notion that there are two options for the normalization constant $\mathcal{N}_\alpha$. These two options, given by $\alpha = 0,1$, correspond to the image-plane normalization (IN) and object-plane normalization (ON), respectively. The value of the normalization in these two cases are given by
\begin{align}
    \mathcal{N}_0 &= \iiint_{-\infty}^\infty \Gamma(u,v) \psi(x+u)\psi^*(x+v)\, \ud u \, \ud v \, \ud x, \label{INval}\\
    \mathcal{N}_1 &= 1. \label{ONval}
\end{align}
Equations~(\ref{INval}) and (\ref{ONval}) show that $\hat{\rho}_0$ always has unit trace, whereas $\hat{\rho}_1$ may not. This distinction is rooted in whether one chooses to budget per-photon information using the photons arriving at the image plane (IN) or those available at the object plane (ON). These two choices, among others, have been a part of the ongoing debate on which normalization leads to FI calculations that are faithful to the physical processes one claims to be studying. A more complete discussion is found in Appendix~\ref{appB}. In what follows, we present the quantum FI calculations for both choices, highlighting the differences between them. One should be clear, however, that the object intensity distribution $I(u) = \Gamma(u,u)$ is normalized to unity in both IN and ON schemes. That is, $\int I(u) \, \ud u = 1$; this follow upon the consideration that the initial state, prior to the dynamic process that maps a state from object to image planes, should have unit trace. From Eqs.~(\ref{INval}) and (\ref{ONval}), it is clear that the choice of $\alpha$ can have non-trivial effects on quantum FI calculations. However, as we will see in Section~\ref{theory}\ref{results}, the derived \textit{asymptotic} FI behaviors (in the sub-Rayleigh regime) have a simple dependence on $\alpha$.

The goal of characterizing a general object scene with arbitrary coherence is taken to be the recovery of $\Gamma$. One may parametrize $\Gamma$ through its (complex) joint moments $\xi_{qp}$, given by
\begin{align}
    \xi_{qp}\triangleq \iint_{-\infty}^\infty \Gamma(u,v) u^q v^p \, \ud u \, \ud v. \label{ximoments}
\end{align}
Under easily-satisfied conditions (although the reconstruction itself may not be trivial), it is known that integrable functions correspond uniquely to their moment sequences. Indeed, for the case of an incoherent object scene, Tsang performs a quantum FI analysis on the moments, $\theta_q$, of $I(u,u) = \Gamma(u,u)$, defined as
\begin{align}
    \theta_q &\triangleq \int_{-\infty}^\infty I(u) u^q \, \ud, \label{thetamoments}
\end{align}
using the Cholesky decomposition to obtain Kraus operators \cite{Tsang2019}.  

Unfortunately, such methods do not extend in the case partially coherent object scenes. This is because $\xi_{qp}$, unlike $\theta_{q+p}$, is not positive definite (although it is still Hermitian) and therefore does not lend itself to a Cholesky decomposition. Nevertheless, one may obtain an alternative Kraus representation using the CMD. The CMD states that every $\Gamma(u,v)$ can be written uniquely as
\begin{align}
    \Gamma(u,v) = \sum_{n=0}^\infty \varphi_n(u)\varphi^*_n(v), \label{CMD}
\end{align}
where $\varphi_n(u)$ are the (modified) coherent modes of $\Gamma(u,v)$. Note that the CMD is more commonly expressed with $\varphi_n(u) = \lambda_n^{1/2}\phi_n(u)$, where $\lambda_n$ is the (positive) modal weight of $\phi_n(u)$, the (unmodified) $n$-th coherent mode of $\Gamma(u,v)$. However, for our purposes, it is more convenient to work with $\varphi_n$. To see how the use of Eq.~(\ref{CMD}) leads to a Kraus representation of $\hat{\rho}_\alpha$, one inserts Eq.~(\ref{CMD}) into Eq.~(\ref{densMat}) and expands the exponentials with their Taylor series to ultimately obtain
\begin{align}
    \hat{\rho}_\alpha &= \mathcal{N}_\alpha^{-1} \sum_{n=0}^\infty \sum_{q,p=0}^\infty \varphi_{qn} \varphi_{pn}^* \frac{(-\ii \hat{k})^q}{q!} |\psi\rangle \langle \psi | \frac{(\ii \hat{k})^p}{p!}, \label{densMatpreK}
\end{align}
where
\begin{align}
    \varphi_{qn} &\triangleq \int_{-\infty}^\infty \varphi_n(u) u^q\,\ud u, \label{cohmodemoments}
\end{align}
is the (complex) $q$-th moment of $\varphi_n(u)$. Equation~(\ref{densMatpreK}) can now be readily written in a Kraus representation as
\begin{align}
    \hat{\rho}_\alpha &= \sum_{n=0}^\infty \hat{A}_{\alpha,n} |\psi\rangle \langle \psi | \hat{A}_{\alpha,n}^\dagger, \label{densMatKraus}
\end{align}
with the Kraus operators $\hat{A}_{\alpha,n}$ given by
\begin{align}
    \hat{A}_{\alpha,n} &\triangleq \mathcal{N}_\alpha^{-1/2} \sum_{q=0}^\infty \varphi_{qn} \frac{(-\ii \hat{k})^q}{q!}. \label{KrausOp}
\end{align}
Note that the Kraus operators in Eq.~(\ref{densMatKraus}) are indexed naturally with the CMD seen in Eq.~(\ref{CMD}). In the following, it is convenient to choose parametrizations of $\Gamma$ that appear explicitly in Eq.~(\ref{KrausOp}). Therefore, we identify the set of real unknown parameters to be $\mathcal{P} = \{\text{Re}(\varphi_{qn}), \text{Im}(\varphi_{qn})\}$ for all integers $n,q \ge 0$. Although $\varphi_{qn} \in \mathbb{C}$, we will often simply write $\varphi_{qn} \in \mathcal{P}$ to refer to either its real or imaginary parts. 

\begin{figure}
    \centering
    \includegraphics[scale=0.52]{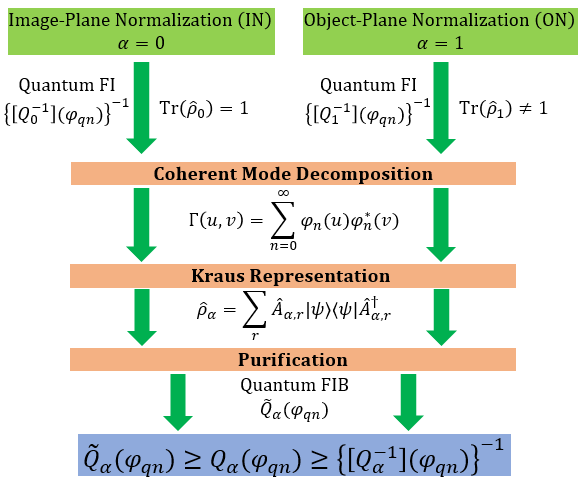}
    \caption{Starting from either IN or ON schemes, the CMD is used to obtain a Kraus representation for $\hat{\rho}_\alpha$. From this form, purification is performed to derive the quantum FIB, $\tilde{Q}_{\alpha}(\varphi_{qn})$, for $\varphi_{qn} \in \mathcal{P}$, the set of unknown parameters. The quantum FIB is an upperbound to the quantum FI.}
    \label{flowchart}
\end{figure}

The standard procedure now is to find the elements of the quantum FI matrix, $Q_{\alpha,ij}$, where $i,j \in \mathcal{P}$. Once this is obtained, the quantum FI for parameter $i$ may be calculated as $([Q_\alpha^{-1}]_{ii})^{-1}$. One may also consider simply calculating $Q_{\alpha,ii}$ since $Q_{\alpha,ii} \ge ([Q_\alpha^{-1}]_{ii})^{-1}$, which provides an upperbound to the quantum FI, and this avoids the need to take the inverse of an infinitely large matrix. Unfortunately, aside from special cases, a general procedure for calculating $Q_{\alpha,ij}$ is known only for \textit{unitary} dynamic processes whose initial state is \textit{pure}. In our context, the mapping from the object to image planes is not unitary [although Eq.~(\ref{densMatKraus}) may be interpreted as a dynamic process with an inital pure state]. Therefore, an attempt to obtain $Q_{\alpha,ij}$ explicitly is mathematically arduous and perhaps intractable. Instead, through a process known as purification, we will obtain matrix elements $\tilde{Q}_{\alpha,ij}$ whose diagonal elements are such that $\tilde{Q}_{\alpha,ii} \ge Q_{\alpha,ii}$. One then has the chain of upperbounds:
\begin{align}
    \tilde{Q}_{\alpha,ii} \ge Q_{\alpha,ii} \ge ([Q_\alpha^{-1}]_{ii})^{-1}. \label{upperbounds}
\end{align}
The quantity on the left-hand side of Eq.~(\ref{upperbounds}) will be called the quantum Fisher information bound (FIB), and note that the first inequality is tight. Its derivation, which depends on purification and the chosen Kraus representation, is given in Appendix~\ref{appC}. We find that
\begin{align}
    \tilde{Q}_{\alpha,ij} = 4 ( C_{\alpha,ij} + B_{\alpha,i}  B_{\alpha,j}), \label{QFIB}
\end{align}
where
\begin{align}
    C_{\alpha,ij} &\triangleq \sum_{r=0}^\infty \langle \psi| \left(\partial_i \hat{A}_{\alpha,r}^\dagger \right)\left(\partial_j \hat{A}_{\alpha,r} \right) |\psi\rangle ,\label{Cop} \\
    B_{\alpha,i} &\triangleq \sum_{r=0}^\infty \langle \psi |\hat{A}_{\alpha,r}^\dagger \left( \partial_i \hat{A}_{\alpha,r} \right)  |\psi\rangle. \label{Bop}
\end{align}
It is possible to re-express Eqs.~(\ref{Cop}) and (\ref{Bop}) in terms of the matrix $\varphi$, whose elements are given by Eq.~(\ref{cohmodemoments}), as
\begin{align}
    C_{\alpha,ij} &= \Tr \left[\Pi \, \partial_i \left(\mathcal{N}_\alpha^{-1/2} \varphi^\dagger \right) \partial_j\left(\mathcal{N}_\alpha^{-1/2} \varphi \right) \right], \label{Cop2}\\
    B_{\alpha,i} &= \mathcal{N}_\alpha^{-1/2} \Tr \left[\Pi \varphi^\dagger \partial_i \left( \mathcal{N}_\alpha^{-1/2} \varphi \right) \right], \label{Bop2}
\end{align}
where
\begin{align}
    \Pi_{pq} &\triangleq \frac{1}{p!q!} \langle \psi| (\ii \hat{k})^p (-\ii \hat{k})^q |\psi \rangle =  \frac{\ii^{p-q}}{p!q!} \!\!\int_{-\infty}^\infty \!\!|\Psi(k)|^2 k^{p+q}\, \ud k,
\end{align}
is related to the moments of the system's optical transfer function. Note that we will assume that $|\psi(k)|^2$ is even. Equations~(\ref{Cop2}) and (\ref{Bop2}), upon insertion into Eq.~(\ref{QFIB}), show how the quantum FIB for any parameter $\varphi_{qn} \in \mathcal{P}$ may be calculated. For convenience, the diagonal elements of $\tilde{Q}_\alpha$ and $C_\alpha$ are denoted as $\tilde{Q}_\alpha(\varphi_{qn})$ and $C_\alpha(\varphi_{qn})$, respectively. Furthermore, we adopt the notation $B_\alpha(\varphi_{qn}) = B_{\alpha,\varphi_{qn}}$. Figure~\ref{flowchart} provides a summary of the procedure used to obtain the quantum FIBs for $\mathcal{P}$.

Although the theoretical framework presented here concerns the quantum FIB, it is useful to also study the per-photon quantum signal-to-noise ratio (QSNR) for $\varphi_{qn}$, which is defined as
\begin{align}
    \text{QSNR}_\alpha(\varphi_{qn}) \triangleq \varphi_{qn}^2 \tilde{Q}_\alpha(\varphi_{qn}).
\end{align}

\subsection{Quantum FIB for moments of the coherent modes} \label{results}
Suppose that $I(u)$ has a characteristic size of $\Delta$. For instance, $I(u)$ may be a Gaussian function with standard deviation $\Delta$, or any (normalizable) function with finite support of width $\Delta$. We are interested in the asymptotic behavior of $\tilde{Q}_\alpha(\varphi_{qn})$, in the limit $\Delta \rightarrow 0$. Recalling that the object-plane position coordinates have been normalized appropriately, the sub-Rayleigh regime is defined as $\Delta \ll 1$. 

Throughout this work, we use the Bachmann-Landau notation, where big $O$ is defined as
\begin{align}
    f(\Delta) = O[g(\Delta)] \Rightarrow \limsup_{\Delta \rightarrow 0} \frac{|f(\Delta)|}{g(\Delta)} < \infty, \label{bigO}
\end{align}
where the the definition includes the limit superior (which may be interpreted as the least upper bound of the quantity's limit points). One interprets Eq.~(\ref{bigO}) as $|f|$ being bounded above (up to some constant factor) by $g$ as $\Delta \rightarrow 0$. In other words, $f$ grows no faster than $g$. We will also use the asymptotic equivalence symbol $\sim$, which is given by
\begin{align}
    f(\Delta) \sim g(\Delta) \Rightarrow \lim_{\Delta \rightarrow 0} \frac{f(\Delta)}{g(\Delta)} = 1. \label{asympequiv}
\end{align}
Equation~(\ref{asympequiv}) is to be interpreted as meaning that $f$ is on the order of $g$ in the stated limit.

We now claim that, in Eq.~(\ref{QFIB}),
\begin{align}
    \tilde{Q}_\alpha (\varphi_{qn}) \sim 4C_\alpha (\varphi_{qn}), \label{Bgone}
\end{align}
for any $\varphi_{qn} \in \mathcal{P}$. Equation~(\ref{Bgone}) states that the second term within the parentheses in Eq.~(\ref{QFIB}) is dominated by the first asymptotically. A detailed discussion of this is given in Appendix~\ref{appD}. At this point, we also note that Eq.~(\ref{Bgone}) expresses the QFIB in terms of infinite sums through the definition in Eq.~(\ref{Cop2}) and is therefore a useful bound only if the sums converge. This is shown to be the case, for practical imaging systems, in Appendix~\ref{appC}. It remains then, to analyze the asymptotic behavior of $C_\alpha(\varphi_{qn})$. The relevant quantities to study, according to Eq.~(\ref{Cop2}), are then $\Pi, \mathcal{N}_\alpha,\varphi$, and the derivatives of the latter two (with respect to $\mathcal{P}$). Clearly, since $\Pi$ and $\mathcal{N}_1 = 1$ do not depend on $\Delta$, we find that
\begin{align}
    \Pi = O(1), \quad \mathcal{N}_1 &= O(1), \quad \frac{\partial \mathcal{N}_1}{\partial \varphi_{qn}} = 0. \label{seteq1}
\end{align}
Note that $\varphi_{qn} = O(\Delta^{q+1/2})$. This, along with the details shown in Appendix~\ref{appD}, give rise to the following asymptotic relations:
\begin{align}
    \mathcal{N}_0^{-1/2} &= O(\Delta^{-1/2}), \quad \varphi = O(\Delta^{1/2}), \nonumber\\
    \text{and} \quad \frac{\partial \mathcal{N}_0^{-1/2}}{\partial \varphi_{qn}} &= O(\Delta^{-1}), \quad \frac{\partial \varphi}{\partial \varphi_{qn}} = O(1). \label{seteq2}
\end{align}
Equations~(\ref{seteq1}) and (\ref{seteq2}) are used in Eq.~(\ref{Cop2}) to give the following results for the quantum FIB:
\begin{align}
    \tilde{Q}_\alpha(\varphi_{qn}) = O(\Delta^{\alpha - 1}), \label{QFIBresult}
\end{align}
which is independent of both $n$ (coherent mode index) and $q$ (moment). The corresponding QSNR is given by
\begin{align}
    \text{QSNR}_\alpha(\varphi_{qn}) &= O(\Delta^{2q+\alpha}), \label{QSNRresult}
\end{align}
which decreases for all $q$, aside from the $q = 0$ case in the IN scheme. Equation~(\ref{QSNRresult}) indicates that it gets progressively more difficult to accurately estimate the higher moments of any $\varphi_n(u)$.

\subsection{Quantum FIB for joint moments of the mutual coherence} \label{PGamma}
Thus far, the results presented in this work concern $\mathcal{P}$, the set of moments of the (modified) coherent modes $\varphi_n$. Although complete knowledge of $\mathcal{P}$ corresponds to knowledge of $\Gamma(u,v)$ [and also $I(u)$], it is of interest to derive the quantum FIB and QSNR for the joint moments of $\Gamma$ directly. That is, we consider $\mathcal{P}_\Gamma = \{\text{Re}(\xi_{qp}),\text{Im}(\xi_{qp})\}$ to be another set of unknown parameters [see Eq.~(\ref{ximoments})]. As was the case with $\mathcal{P}$, we will often refer to an element in $\mathcal{P}_\Gamma$ as simply $\xi_{qp}$ to mean either its real or imaginary part.

The strategy is to compute the Jacobian matrix
\begin{align}
    J_\Gamma = \frac{\partial \mathcal{P}_\Gamma}{\partial \mathcal{P}},
\end{align}
and use the reparametrization property of quantum FI (bound) matrices:
\begin{align}
    \tilde{Q}_{\alpha,\Gamma}^{-1} = J_\Gamma \cdot  \tilde{Q}_{\alpha}^{-1} \cdot J_\Gamma^\dagger, \label{reparameter}
\end{align}
where we use the notation $\tilde{Q}_{\alpha,\Gamma}^{-1}$ to denote the inverse of the quantum FIB matrix on $\mathcal{P}_\Gamma$. Note that Eq.~(\ref{reparameter}) is given as the inverse of the quantum FIB matrix and so expresses information directly about the Cram\'{e}r-Rao bound (bound) (CRBB) of $\mathcal{P}_\Gamma$ parameters rather than the information. To return to information bounds, we consider inverting the diagonal elements of Eq.~(\ref{reparameter}) to obtain
\begin{align}
    [\tilde{Q}_{\alpha,\Gamma}^{-1} (\mathcal{P}_\Gamma)]^{-1} = \{[J_\Gamma \cdot  \tilde{Q}_{\alpha}^{-1} \cdot J_\Gamma^\dagger](\mathcal{P}_\Gamma)\}^{-1}, \label{effQFIB}
\end{align}
where $M(\mathcal{P}_\Gamma)$ denotes the diagonal elements of a matrix $M$ with respect to parametrization $\mathcal{P}_\Gamma$. Equation~(\ref{effQFIB}) is a means to calculate the quantum FIB for the parameters in $\mathcal{P}_\Gamma$. For clarity, however, we point out that Eq.~(\ref{effQFIB}) is not strictly the same quantity as $\tilde{Q}_\alpha (\mathcal{P})$ calculated in Eq.~(\ref{QFIBresult}). First, Eq.~(\ref{effQFIB}) refers to a new set of parameters $\mathcal{P}_\Gamma$ and, second, it appears in the middle of the following chain of inequalities:
\begin{align}
    \tilde{Q}_{\alpha,\Gamma}(\mathcal{P}_\Gamma) \ge [\tilde{Q}_{\alpha,\Gamma}^{-1} (\mathcal{P}_\Gamma)]^{-1} \ge [Q_{\alpha,\Gamma}^{-1} (\mathcal{P}_\Gamma)]^{-1}. \label{newchain}
\end{align}
That is, the quantity calculated in Eq.~(\ref{QFIBresult}) is analogous to the first term in Eq.~(\ref{newchain}), and the one in Eq.~(\ref{effQFIB}) is found in the middle. However, we are justified in calling Eq.~(\ref{effQFIB}) a QFIB as it is still greater than or equal to the QFI, which is the final term in Eq.~(\ref{newchain}). The need for this clarification comes from the fact that Eq.~(\ref{reparameter}) involves the inverses of the quantum FIB matrices, which was in turn necessitated through the appearance of $J_\Gamma$ in Eq.~(\ref{reparameter}) rather than its inverse. The reason for this choice, as we will see in Eq.~(\ref{xivarphirelate}), is that it is easy to express $\xi_{qp}$ as a function of $\varphi_{qn}$ but the converse is not true. 

To proceed, we note that Eq.~(\ref{effQFIB}) involves $\tilde{Q}_\alpha^{-1}$ the inverse of quantum FIB matrix for $\mathcal{P}$. Using Eqs.~(\ref{seteq2}) and (\ref{QFIBresult}), it is possible to show that 
\begin{align}
    [\tilde{Q}_\alpha]_{(qn)(pm)} = O(\Delta^{\alpha-1}),
\end{align}
where we use the double-subscripts $(qn)$ to refer to the element $\varphi_{qn} \in \mathcal{P}$. That is, every element in $\tilde{Q}_\alpha$ has the same asymptotic dependence on $\Delta$. This allows for a straightforward analysis of the asymptotic behavior of $\tilde{Q}_\alpha^{-1}$. Namely,
\begin{align}
    [\tilde{Q}_{\alpha}^{-1}]_{(qn)(pm)} = O(\Delta^{1-\alpha}). \label{inverseQFIB}
\end{align}

In order to calculate the Jacobian matrix $J_\Gamma$, one must first express $\xi_{qp} \in \mathcal{P}_\Gamma$ as a function of the parameters $\varphi_{rn} \in \mathcal{P}$. By multiplying both sides of Eq.~(\ref{CMD}) by $u^qv^p$ and integrating over $(u,v)$, we find that
\begin{align}
    \xi_{qp} = \sum_{n=0}^\infty \varphi_{qn}\varphi_{pn}^*. \label{xivarphirelate}
\end{align}
Note that, for all $n$, $\xi_{qp}$ is only related to the $q$-th and $p$-th moments of $\varphi_n$. One may now calculate the elements of $J_\Gamma$ using Eq.~(\ref{xivarphirelate}). Henceforth, for convenience, we will take $\varphi_{qn}\in\mathcal{P}$ to mean the \textit{real part} of $\varphi_{qn}$; it is straightforward to see that the corresponding expressions regarding the imaginary parts of $\varphi_{qn}$ are similar. With this,
\begin{align}
    [J_\Gamma]_{(qp)(rn)} = \frac{\partial \xi_{qp}}{\partial \varphi_{rn}} =\delta_{rq}\varphi_{pn}^* + \delta_{rp} \varphi_{qn}, \label{JGammaexp}
\end{align}
where, without loss of generality (due to the Hermiticity of $\xi$), we take $q \ge p$. Equation~(\ref{JGammaexp}) indicates that the elements of $J_\Gamma$ depend linearly on  $\mathcal{P}$ and the final equality uses the asymptotic behavior of $\varphi_{qn}$. Using Eq.~(\ref{effQFIB}), we find that
\begin{align}
    [\tilde{Q}_{\alpha,\Gamma}^{-1} (\xi_{qp})]^{-1} &= \{[J_\Gamma \cdot  \tilde{Q}_{\alpha}^{-1} \cdot J_\Gamma^\dagger](\xi_{qp})\}^{-1} \nonumber\\
    &= [O(\Delta^{p+1/2}) O(\Delta^{1-\alpha}) O(\Delta^{p+1/2})]^{-1} \nonumber\\
    &=O(\Delta^{-2p+\alpha-2}). \label{QFIBxi}
\end{align}
Note that the second line in Eq.~(\ref{QFIBxi}) arises from Eq.~(\ref{JGammaexp}) and noting that the $r = q$ term gives the dominant asymptotic behavior $O(\Delta^{p+1/2})$. Using $\xi_{qp} = O(\Delta^{q+p+1})$, we can derive a QSNR for $\xi_{qp}$ as
\begin{align}
    \text{QSNR}_{\alpha,\Gamma}(\xi_{qp}) &\triangleq \xi_{qp}^2 [\tilde{Q}_{\alpha,\Gamma}^{-1} (\xi_{qp})]^{-1} = O(\Delta^{2q+\alpha}). \label{QSNRxi}
\end{align}

\subsection{Quantum FIB for the moments of the intensity distribution}\label{PI}
For the purposes of \textit{imaging} the object scene, one may only be interested in $I(u)$ rather than the entire mutual coherence function $\Gamma(u,v)$. In this case we may consider $\mathcal{P}_I = \{\theta_q\}$, the moments of $I(u)$, defined in Eq.~(\ref{thetamoments}). 

It is shown in Appendix~\ref{appF} that
\begin{align}
    \theta_q &= \frac{1}{(2\pi)^q}\sum_{n=0}^\infty \left\{\! \int_{-\infty}^\infty \!\!\left[ \sum_{k_1,k_2 = \lceil q/2 \rceil}^\infty \!\!\!\! A_q^{k_1 k_2}(f) \varphi_{k_1 n} \varphi_{k_2 n}^* \right] \!\! \ud f \! \right\}, \label{varphitotheta}
\end{align}
where $A_q^{k_1 k_2}(f)$ is some $\Delta$-independent quantity. Crucially, $\theta_q$ depends, for all coherent modes $n$, only on (products of) moments of $\varphi_n$ that are at least as large as $\lceil q/2 \rceil$. Now, the desired Jacobian 
\begin{align}
    J_I = \frac{\partial \mathcal{P}_I}{\partial \mathcal{P}},
\end{align}
has elements $[J_I]_{q(rn)} = \partial \theta_q / \partial \varphi_{rn}$ given by
\begin{align}
    [J_I]_{q(rn)} \! &=\! \frac{1}{(2\pi)^q} \! \sum_{n=0}^\infty \! \left\{ \! \int_{-\infty}^\infty \!\!\left[ \sum_{k=\lceil q/2 \rceil}^\infty  \!\!\!\!2\text{Re}\{A^{k r}_q (f) \varphi_{kn}\}   \right] \!\! \ud f \! \right\}, \label{jacobianI}
\end{align}
where we used $A^{k_1 k_2}_q = (A^{k_2 k_1}_q)^*$. Using Eq.~(\ref{effQFIB}), with $\Gamma$ replaced with $I$, we find that
\begin{align}
    &[\tilde{Q}_{\alpha,I}^{-1} (\theta_q)]^{-1} = \{[J_I \cdot  \tilde{Q}_{\alpha}^{-1} \cdot J_I^\dagger](\xi_{qp})\}^{-1} \nonumber\\
    &= \left[\sum_{k_1,k_2=\lceil q/2 \rceil}^\infty O(\Delta^{k_1+1/2}) O(\Delta^{1-\alpha}) O(\Delta^{k_2+1/2}) \right]^{-1} \nonumber\\
    &=O(\Delta^{-2\lceil q/2 \rceil+\alpha-2}). \label{QFIBtheta}
\end{align}
The second line in Eq.~(\ref{QFIBtheta}) arises from Eq.~(\ref{jacobianI}) and noting that the $k=\lceil q/2 \rceil$ term gives the dominant asymptotic behavior $O(\Delta^{\lceil q/2 \rceil + 1/2})$ (for both $k_1,k_2$ summations). Using $\theta_q = O(\Delta^q)$, one may now derive a QSNR for $\theta_q$ as
\begin{align}
    \text{QSNR}_{\alpha,I}(\theta_q) &\triangleq \theta_q^2 [\tilde{Q}_{\alpha,I}^{-1} (\theta_q)]^{-1} = O(\Delta^{2\lfloor q/2 \rfloor +\alpha-2}), \label{QSNRtheta}
\end{align}
which is valid for all $q$.

\section{Results and Discussion}

\begin{figure}[ht]
    \centering
    \includegraphics[scale=0.65]{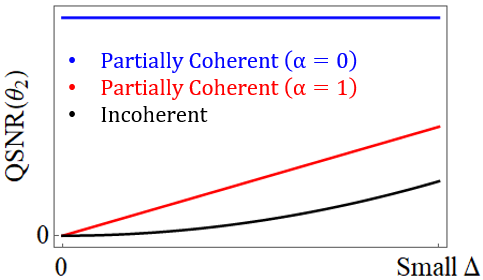}
    \caption{Plot that shows the qualitative behavior, in the sub-Rayleigh regime, of $\text{QSNR}(\theta_2)$. The results derived here for the partially coherent case (blue and red) both fall off asymptotically slower than the incoherent QSNR derived in [Tsang].}
    \label{exampletheta2}
\end{figure}

The quantum FIB and QSNR results for the parameters in the considered sets of $\mathcal{P},\mathcal{P}_\Gamma,$ and $\mathcal{P}_I$ are shown in Table~\ref{summarytable}. There are several remarks to be made:
\begin{itemize}
    \item The coherent mode index (often denoted as $n$ or $m$ throughout this work) does not appear in either the quantum FIB or QSNR for $\varphi_{qn}\in \mathcal{P}$. This means that the quantum bounds we derive here are independent on which coherent mode's moments are being considered.
    \item The quantum FIB for $\varphi_{qn}\in \mathcal{P}$ is independent of $q$, but the corresponding QSNR vanishes as $\Delta \rightarrow 0$ (aside from $q=0$ and $\alpha = 0$, for which the QSNR approaches a constant value asymptotically).
    \item The QSNR for $\theta_q \in \mathcal{P}_I$, given by $O(\Delta^{2\lfloor q/2 \rfloor+\alpha -2})$, vanishes more slowly in the limit $\Delta \rightarrow 0$ when compared to the incoherent result reported in Tsang, in which $\text{QSNR} (\theta_q) = O(\Delta^{2\lceil q/2 \rceil})$ \cite{Tsang2019}. This is true for both IN and ON schemes. In fact, for the ON scheme, we note that the the $\text{QSNR}_{1,I}(\theta_2) = O(1)$ which is non-zero as $\Delta \rightarrow 0$. This is shown in Fig.~\ref{exampletheta2}
\end{itemize}

\begin{table}[ht]
\caption{\label{summarytable}%
Summary of the asymptotic behavior ($\Delta \rightarrow 0$) of the quantum FIB and QSNR for the various moments explored in this work. The IN and ON schemes correspond to $\alpha = 0,1$, respectively. Note that $q \ge p$ for $\xi_{qp}$. 
}
\begin{ruledtabular}
\begin{tabular}{lcc}
\multicolumn{1}{c}{\textrm{Parameters}}&
Quantum FIB & QSNR\\
\colrule
$\varphi_{qn} \in \mathcal{P}$ & $O(\Delta^{\alpha - 1})$ & $O(\Delta^{2q+\alpha})$ \\
$\xi_{qp} \in \mathcal{P}_\Gamma$ & $O(\Delta^{-2p + \alpha -2})$ & $O(\Delta^{2q+\alpha})$ \\
$\theta_q \in \mathcal{P}_I$  & $O(\Delta^{-2\lceil q/2 \rceil +\alpha - 2})$ & $O(\Delta^{2\lfloor q/2 \rfloor+\alpha -2})$ \\
\end{tabular}
\end{ruledtabular}
\end{table}

Complete knowledge of $\mathcal{P}$ or $\mathcal{P}_\Gamma$ allows one to accurately reconstruct $\Gamma(u,v)$, the entire object mutual coherence function. The results summarized in Table~\ref{summarytable} indicate that the QSNR for these parameters decay for higher moments, which implies the increasing difficulty in their estimation. On the other hand, when only the intensity information of the object is of interest, one might instead consider $\mathcal{P}_I$. The QSNR for $\theta_q$ is shown to fall off more slowly when compared to the incoherent case. The possibility of estimating the $I(u)$ from a partially coherent object more accurately than one could from an incoherent object thereby remains open from these quantum bound results.

Care should be taken in interpreting these results, however, since quantum FIB calculations do not explicitly instruct on how the optimal measurement (if it exists) can be found. Furthermore, even if we are certain that such a measurement exists, we are still at the moment ignorant to what families of $\Gamma(u,v)$ would lend to a saturation of the results in Table~\ref{summarytable}. In the best case scenario, the derived bounds are saturated for arbitrary $I(u)$ and a particular family of $\gamma(u,v)$, where $\gamma(u,v) \triangleq \Gamma(u,v) [I(u)I(v)]^{-1/2}$ is the complex degree of coherence. Here, one may envision the possibility of precise sub-Rayleigh imaging/communication as long as one properly chooses $\gamma(u,v)$. However, if the derived bounds correspond to particular distributions of $I(u)$, then the possibilities are drastically restricted. 

Finally, we discuss the difference between the IN and ON schemes, which correspond to scenarios that are detailed in Appendix~\ref{appB}. Although it was not immediately obvious upon inspecting the definitions of $\mathcal{N}_\alpha$, the analysis in this manuscript demonstrates that the difference between the IN and ON schemes with regards to the resulting quantum bounds (in the asymptotic limit) may be summarized as
\begin{align}
    \frac{\tilde{Q}_0(x)}{\tilde{Q}_1(x)} = \frac{\text{QSNR}_0(x)}{\text{QSNR}_1(x)} = O(\Delta^{-1}), \label{quantumRatio}
\end{align}
where $x$ is any parameter in $\mathcal{P},\mathcal{P}_\Gamma,$ or $\mathcal{P}_I$. Equation~(\ref{quantumRatio}) shows that the IN scheme gives a quantum FIB and QSNR that grows more quickly (or decays more slowly) as $\Delta \rightarrow 0$ when compared to the ON scheme. However, in the case of an incoherent object scene, Appendix~\ref{appB} shows that there is no distinction between IN and ON because $\mathcal{N}_\alpha$ is independent of the unknown parameters.

\section{Concluding Remarks}
A framework for calculating quantum information bounds for estimating parameters from a general, one-dimensional, partially coherent object with mutual coherence $\Gamma$ is introduced. Using the CMD, purification, and reparametrization, we find the QFIB and QSNR for various object moments in both IN and ON normalization schemes. 

There are many possible generalizations and extensions of the results presented here. For instance, an analysis of the higher spatial dimension case remains open although analogous results are expected there. A stricter (non-asymptotic) calculation of the QFIB (or the QFI itself) beyond purification and asymptotics is also desirable since, as seen in Ref.~\onlinecite{Liang2021}, the quantum FI can vary quickly within the sub-Rayleigh regime and the big $O$ results may fail to completely capture this behavior. Finally, it is prudent to see if the derived QFIBs can be attained experimentally. A first step in this exploration is to compute estimators for object moments under SPADE-like measurement schemes which, for incoherent objects, have been shown to be optimal.

\begin{acknowledgments}
We acknowledge support from the Defense Advanced Research Projects Agency (D19AP00042)
\end{acknowledgments}

\appendix

\section{Density Matrix for a General Partially Coherent Object Distribution} \label{appA}
Following the derivation in the Supplemental Material of Ref.~\onlinecite{Liang2021}, one obtains that the density matrix representing one-photon detection events from a multimode object distribution is given by
\begin{align}
    \hat{\rho} = \frac{1}{\epsilon} \sum_{j,k=1}^M  G_{jk} |1_j\rangle\langle 1_k|, \label{densMat1}
\end{align}
where $G_{jk}$ are the matrix elements $\hat{G}$, the image-plane mutual coherence matrix and $\epsilon$ is the average number of photons detected on the image plane over a coherence time. Note that $|1_j\rangle$ is the 1-photon state in the $j$-th mode, of which there are $M$ total. To proceed, we relate $\hat{G}$ to $\hat{\Gamma}$, the object-plane mutual coherence matrix, through the scattering equation
\begin{align}
    \hat{G} = \hat{S} \hat{\Gamma} \hat{S}^\dagger,
\end{align}
where $\hat{S}$ is the (imaging) system's field scattering matrix. Often, $\hat{S}$ is not unitary and is closely related to the system's point spread function (PSF). For general object coherence, the matrix elements of $\hat{\Gamma}$ will be indicated by $[\hat{\Gamma}]_{uv} = \epsilon_0 \Gamma_{uv}$, where $\epsilon_0$ is the average photon number emitted from the source over the coherence time of the source. Equation~(\ref{densMat1}) can now be expressed as
\begin{align}
    \hat{\rho} &\approx \frac{\epsilon_0}{\epsilon} \sum_{u,v} \Gamma_{uv}\sum_{j,k=1}^M S_{ju} S^*_{kv} |1_j\rangle \langle 1_k |. \label{densMat2}
\end{align}
At this point, we stipulate that $\epsilon = \epsilon_0 \eta \mathcal{N}_\alpha$, where $\eta$ is the quantum efficiency of the system and we assume that this is the same for every object mode. $\mathcal{N}_\alpha$ is a normalization constant that could take two possible values $(\alpha = 1,2)$ that correspond to two interpretations of the density matrix, which will now similarly be indexed as $\hat{\rho}_\alpha$. The meaning of $\alpha = 1,2$ will be discussed in Appendix~\ref{appC}. 

By defining
\begin{align}
    |\psi_u \rangle = \sum_{j=1}^M \psi(j,u) |1_j\rangle \quad \text{with} \quad \psi(j,u) \triangleq S_{ju}/\sqrt{\eta},
\end{align}
one may write Eq.~(\ref{densMat2}) as
\begin{align}
    \hat{\rho}_\alpha &= \mathcal{N}_\alpha^{-1} \sum_{u,v} \Gamma_{uv}\sum_{j,k=1}^M \psi(j,u) \psi^*(k,v) |1_j\rangle \langle 1_k |. \label{densMat3}
\end{align}
At this point, the paraxial approximation is used to allow image-plane wavepacket positions to be used to represent the image plane modes. That is, one identifies $x_j = x_0 + j \ud x$ and $|x_j\rangle = |1_j \rangle/\sqrt{\ud x}$, where $x_0$ is an arbitrary choice of origin and $\ud x$ is a the spacing between image-plane positions. By further rewriting $\psi_u(x_j) = \psi(j,u)/\sqrt{\ud x}$, it is possible to obtain the succinct equation of
\begin{align}
    \hat{\rho}_\alpha &= \mathcal{N}_\alpha^{-1} \sum_{u,v} \Gamma_{uv} |\psi_u\rangle \langle \psi_v|, \label{densMat4}
\end{align}
where
\begin{align}
    |\psi_u\rangle \triangleq \sum_{j=1}^M \psi_u (x_j) |x_j\rangle\,\ud x.
\end{align}
In the (image-plane) continuum limit of $\ud x\rightarrow 0$, one obtains that $|\psi_u\rangle \rightarrow \int \psi_u(x) |x\rangle \, \ud x$. For the special case where the system is shift-invariant, 
\begin{align}
    |\psi_u \rangle = \int_{-\infty}^\infty \psi_u(x) |x\rangle \,\ud x = e^{-\ii \hat{k} u} \int_{-\infty}^\infty \psi(x) |x\rangle \,\ud x,
\end{align}
where $\psi(x)$ is the shift-invariant PSF. In all, one may now write
\begin{align}
    \hat{\rho}_\alpha &= \mathcal{N}_\alpha^{-1} \sum_{u,v} \Gamma_{uv} e^{-\ii \hat{k}u} |\psi \rangle \langle \psi| e^{\ii \hat{k}v}.
\end{align}
The final step is to invoke the position-representation and continuum limit for the object plane coordinates. This can be summarized by taking $\Gamma_{uv} = \Gamma(u,v)\, \ud u \, \ud v$, where $\Gamma(u,v)$ is the familiar mutual coherence function that one thinks of in classical partial coherence theory and $\ud u$ or $\ud v$ are spacings in the object-plane positions. The continuum limit then corresponds to $(\ud u,\ud v) \rightarrow (0,0)$, resulting in
\begin{align}
    \hat{\rho}_\alpha = \mathcal{N}_\alpha^{-1} \iint_{-\infty}^\infty \Gamma(u,v) e^{-\ii \hat{k} u} |\psi \rangle \langle \psi | e^{\ii \hat{k} v} \, \ud u \, \ud v. \label{densMatAppendix}
\end{align}

\section{Normalization of the density matrix} \label{appB}

This section concerns the values of $\mathcal{N}_0$ and $\mathcal{N}_1$, which are the normalization factors for the image-plane normalization (IN) and object-plane normalization (ON) schemes, respectively. These two schemes correspond to two different assumptions regarding the (mixed) state at the measurement stage. For both schemes, the initial state (object plane distribution) is normalized so that
\begin{align}
    \int_{-\infty}^\infty \Gamma(u,u) \, \ud u = \int_{-\infty}^\infty I(u) \, \ud u = 1. \label{initialnormal}
\end{align}
Equation~(\ref{initialnormal}) simply reflects the notion that the density matrix at the object plane must be normalized so that, when projected on an orthonormal basis, the resulting expansion coefficients may be interpreted as probability densities. However, in general imaging systems, the transfer of state between the object and image plane is not unitary (finite apertures) and therefore does not necessarily preserve the unit trace property of the initial state.

The IN scheme is the one in which the image-plane density matrix is renormalized after ungoing the non-unitary process, giving $\hat{\rho}_0$ such that $\Tr(\hat{\rho}_0) = 1$. Although the quantum FI informs on a measurement-free upperbound on the precision of estimating a parameter, one can liken the IN scheme to one where the experimentalist chooses to normalize the photon counts of each measurement used to perform maximum likelihood estimation (MLE). On the otherhand, the ON scheme may be interpreted as the procedure where the experimentalist does not normalize their measurements (as they are successively performed for MLE) by the photon number. Therefore, one simply takes $\mathcal{N}_1 = 1$.

\subsection{Calculating normalization for IN}

The value for $\mathcal{N}_0$ is calculated, using the orthonormal set of position kets $\{|x\rangle\}$ to perform the trace, to be
\begin{align}
    &\mathcal{N}_0 = \Tr \left[ \iint_{-\infty}^\infty \Gamma(u,v) e^{-\ii \hat{k} u} |\psi \rangle \langle \psi | e^{\ii \hat{k} v} \, \ud u \, \ud v \right] \nonumber \\
    &= \int_{-\infty}^\infty \langle x |\left[ \iint_{-\infty}^\infty \Gamma(u,v) e^{-\ii \hat{k} u} |\psi \rangle \langle \psi | e^{\ii \hat{k} v} \, \ud u \, \ud v \right] | x \rangle \, \ud x \nonumber \\
    &= \iiint_{-\infty}^\infty \Gamma(u,v) \psi(x+u) \psi^*(x+v) \, \ud u \, \ud v \, \ud x. \label{n1value}
\end{align}
Because $\mathcal{N}_0$ contains $\Gamma$, it is clear that it depends on the set of parameters to be estimated $\mathcal{P} = \{\lambda_n, \varphi_{n,q}\}$, which refer to the coherent mode decomposition (CMD) of $\Gamma$:
\begin{align}
    \Gamma(u,v) &= \sum_{n=0}^\infty \lambda_n \phi_n(u) \phi_n^*(v), \label{CMDapp}\\
    \text{with} \quad \varphi_{qn} &\triangleq \int_{-\infty}^{\infty}\lambda_n^{1/2}\phi_n(u) u^q \, \ud u.
\end{align}
To see the dependence of $\mathcal{N}_0$ on $\mathcal{P}$ explicitly, we expand the PSF quantities in Eq.~(\ref{n1value}) in a Taylor series and insert Eq.~(\ref{CMDapp}) into the resulting expression. This ultimately results in
\begin{align}
    \mathcal{N}_0 &= \sum_{n=0}^\infty \sum_{q,p =0}^\infty \frac{H_{qp}}{q!p!} \varphi_{qn} \varphi_{pn}^*, \label{n1valexp}
\end{align}
where
\begin{align}
    H_{qp} &\triangleq \int [\partial_x^{(q)}\psi(x)] [\partial_x^{(p)}\psi^*(x)] \, \ud x,
\end{align}
and $\partial_x^{(q)}$ denotes the $q$-th derivative w.r.t $x$. Note that, for an real and even PSF (e.g. a Gaussian function), $H_{qp}$ vanishes when $q + p \in 2\mathbb{Z} + 1$. Equation~(\ref{n1valexp}) shows the explicit dependence of $\mathcal{N}_1$ on $\mathcal{P}$. 

Note that in the case of an incoherent object distribution,  $\Gamma(u,v) = \delta(u-v)I(u)$, and Eq.~(\ref{n1value}) becomes
\begin{align}
    \mathcal{N}_0 = \left[ \int_{-\infty}^\infty I(u) \, \ud u \right] \left[ \int_{-\infty}^\infty |\psi(x)|^2 \, \ud x \right] = 1,
\end{align}
if one normalizes the incoherent PSF to unity (otherwise, $\mathcal{N}_0$ is still another constant). This result shows that there is no distinction between the IN and ON schemes. This observation has been observed and detailed before in the analyses regarding two-point separation estimation.

\section{Purification and the quantum FI bound} \label{appC}

For any initial pure state, $|\psi\rangle \langle \psi |$, any dynamical process that results in the final (possibly mixed) state $\hat{\rho}$ can be represented with Choi-Kraus operators $\hat{A}_r$ such that
\begin{align}
    \hat{\rho}(\{x\}) = \sum_{r=0}^\infty \hat{A}_r(\{x\}) |\psi\rangle \langle \psi| \hat{A}_r^\dagger(\{x\}), \label{dynamicprocess}
\end{align}
where we show the dependence of $\hat{\rho}$ and $\hat{A}$ on the set of unknown parameters $\{x\}$. Note that the Choi-Kraus operators satisfy $\sum_r \hat{A}_r \hat{A}^\dagger_r = \hat{I}$, where $\hat{I}$ is the identity operator. However, the transformation represented in Eq.~(\ref{dynamicprocess}) does not need to be unitary. When the dynamical process is not unitary (like in imaging systems) or when $\hat{\rho}$ is not pure, there is no general procedure to calculate the quantum FI for the parameters $\{x\}$, denoted $Q_i$ for $x_i \in \{x\}$. However, it is possible to perform a process termed purification in order to find a quantum FI bound (FIB), denoted $\tilde{Q}_i$ for $x_i \in \{x\}$.

In purification, one denotes the Hilbert space of $|\psi\rangle \langle \psi|$ and $\hat{\rho}$ as $S$, for the system. An auxiliary environmental Hilbert space, denoted $E$, equipped with an orthonormal set of states $\{|r\rangle_E\}$ is introduced. The motivation of purification is to allow $\hat{\rho} \in S$ to be seen as a pure state in the combined Hilbert space $S+E$. To do this, one introduces a pure state $|\Psi(\{x\})\rangle$ in $S+E$ and a unitary operator $\hat{U}(\{x\}) \in S+E$  such that 
\begin{align}
    |\Psi(\{x\})\rangle = \hat{U}(\{x\}) |\psi\rangle_S |0\rangle_E \triangleq \sum_r \hat{A}_r(\{x\}) |\psi\rangle_S |r\rangle_E, \label{intropure}
\end{align}
where we use the subscript $S$ to emphasize the that $|\psi\rangle$ is a state of the system only. Using the orthonormality of $\{|r\rangle_E\}$, Eq.~(\ref{intropure}) can be used to relate
\begin{align}
    \hat{A}_r(\{x\}) = \langle r|_E \hat{U}(\{x\})|0\rangle_E.
\end{align}
The purification process of $\hat{\rho}(\{x\}) \rightarrow |\Psi(\{x\})\rangle$ should be checked to ensure that when $E$ is traced out, one recovers the original mixed state in $S$. Indeed one may verify immediately from Eq.~(\ref{intropure}) that
\begin{align}
    \Tr_E[|\Psi(\{x\})\rangle\langle \Psi(\{x\})|] = \hat{\rho}(\{x\}).
\end{align}

To recapitulate, purification is done so that, in the space $S+E$, the transformation in Eq.~(\ref{dynamicprocess}) is unitary and one may perform quantum FI calculations on $|\Psi(\{x\})\rangle \langle \Psi(\{x\})|$ (a pure state) instead of on $\hat{\rho}(\{x\})$. That is, we may use the result for quantum FI procedures on pure states to state that the quantum FI matrix for $|\Psi(\{x\})\rangle \langle \Psi(\{x\})|$, denoted $\tilde{Q}$, is given by
\begin{align}
    \tilde{Q}_{ij} &= 4 \left( \langle \Psi| \hat{K}_i \hat{K}_j |\Psi\rangle - \langle \Psi| \hat{K}_i |\Psi\rangle \langle \Psi| \hat{K}_j | \Psi\rangle \right), \label{QFIBorigins}
\end{align}
where
\begin{align}
    \hat{K}_i \triangleq \ii \left(\frac{\partial \hat{U}}{\partial x_i} \right) \hat{U}^\dagger, \label{generator}
\end{align}
is the Hermitian generator of displacements for the parameter $x_i$. Note that the explicit dependency of quantities on $\{x\}$ has been omitted for convenience. Using Eqs.~(\ref{intropure}) and (\ref{generator}) in Eq.~(\ref{QFIBorigins}), we find that it is possible to express $\tilde{Q}_{ij}$ in an $E$-independent manner:
\begin{align}
    \tilde{Q}_{ij} &= 4 \left( \langle \psi| C_{ij} |\psi \rangle + \langle \psi| B_i |\psi\rangle \langle \psi| B_j | \psi \rangle  \right),
\end{align}
where
\begin{align}
    C_{ij} &\triangleq \sum_r \langle \psi| \left(\partial_i \hat{A}^\dagger_r \right) \left( \partial_j \hat{A}_r \right) |\psi\rangle,\\
    B_i &\triangleq \sum_r \langle \psi| \hat{A}_r^\dagger\left(\partial_i \hat{A}_r\right)  |\psi \rangle.
\end{align}

It is important to emphasize that $\tilde{Q}$ is the quantum FI matrix for $|\Psi\rangle \langle \Psi|$ and not $Q$, the quantum FI matrix for $\hat{\rho}$. However, one can show that $\tilde{Q}_{ii} \ge Q_{ii}$, which is the reason why $\tilde{Q}_{ii}$ is known as the quantum FI \textit{bound} for the parameter $x_i$. To see this, we introduce $\{\hat{E}^S_j\}$ as a set of positive operator-valued measures (POVM) over the Hilbert space $S$. For a given POVM, the classical FI for $x_i$ is denoted as $F_{ii}(\{\hat{E}^S_j\})$. Then,
\begin{align}
    Q_{ii} &= \max_{\{\hat{E}^S_j\}} F_{ii}(\{\hat{E}^S_j\}) = \max_{\{\hat{E}^S_j\}} F_{ii}(\{\hat{E}^S_j\} \otimes \hat{I}^E) \nonumber\\
    &\le \max_{\{\hat{E}^{S+E}_j\}} F_{ii}(\{\hat{E}_j^{S+E}\}) = \tilde{Q}_{ii}, \label{boundchain}
\end{align}
which gives us the desired inequality. The inequality arises from the notion that $\{\hat{E}^S_j\} \otimes \hat{I}^E$ is just one of infinitely many POVMs in $S+E$, denoted $\{\hat{E}_j^{S+E}\}$. Furthermore, as seen in Ref.~\onlinecite{escher}, the inequality in Eq.~(\ref{boundchain}) is tight.

\subsection{Convergence of the quantum FI bound}
The following asymptotic relation holds true, as shown in Appendix~\ref{appD}, regardless of normalization scheme:
\begin{align}
    \tilde{Q}_{\alpha,\varphi_{rn}\varphi_{pm}} \sim 4 \mathcal{N}_\alpha^{-1} \Tr \left[ \Pi \frac{\partial \varphi}{\partial \varphi_{rn}} \frac{\partial \varphi}{\partial \varphi_{pm}} \right], \label{QFIBasymp}
\end{align}
for $\varphi_{rn},\varphi_{pm} \in \mathcal{P}$. $\tilde{Q}_\alpha$ is positive-semidefinite and therefore,
\begin{align}
    |\tilde{Q}_{\alpha,\varphi_{rn}\varphi_{pm}}| \le \sqrt{\tilde{Q}_{\alpha}(\varphi_{rn}) \tilde{Q}_{\alpha}(\varphi_{pm})}, \label{positivesemi}
\end{align}
Equation~(\ref{positivesemi}) shows that, in order to prove that $\tilde{Q}_{\alpha,\varphi_{rn}\varphi_{pm}}$ is bounded, it is sufficient to show that $\tilde{Q}_{\alpha,\varphi_{pn}}$ is.

From Eq.~(\ref{QFIBasymp}), one may show that
\begin{align}
    \tilde{Q}_{\alpha,\varphi_{rn}\varphi_{pm}} \sim 4 \mathcal{N}_\alpha^{-1}\Tr \left[ \Pi \frac{\partial \varphi}{\partial \varphi_{rn}} \frac{\partial \varphi}{\partial \varphi_{pm}} \right] \le 4 \|\Pi\| \left\| \partial_i \varphi \right\|^2_2, \label{norms}
\end{align}
where $\|\cdot\|$ is the operator norm of $\Pi$ and $\|\cdot \|_2$ is the Hilbert-Schmidt ($L^2$) norm. Therefore, it is sufficient to show $\tilde{Q}_{ii}$ is bounded by showing that both $\|\Pi\|$ and $\|\partial_i \varphi\|_2$ are bounded. 

It is shown in Ref.~\onlinecite{Tsang2019} that, among other cases, $\Pi$ is bounded in the operator norm if the OTF of the system is bandlimited. It remains to be show tnat $\partial \varphi/\partial \varphi_{pm}$ is bounded in the Hilbert-Schmidt norm for all $\varphi_{pm} \in \mathcal{P}$. But note that
\begin{align}
    \frac{\partial \varphi_{qn}}{\partial \varphi_{pm}} = \delta_{nm} \delta_{qp},
\end{align}
and so $\partial_{\varphi_{pm}} \varphi$ is non-zero only in the matrix element $(p,m)$. $\|\partial \varphi /\partial \varphi_{pn}\|^2 = 1 < \infty$, and so the desired convergence is shown.

\section{Asymptotic behavior of various quantities} \label{appD}

Note that, using the CMD,
\begin{align}
    \Gamma(u,v) = \sum_{n=0}^\infty \varphi_n(u) \varphi_n^*(v).
\end{align}
Therefore, since $\Gamma(u,v) \propto (\Delta^{-1})$ [due to the normalization of $I(u) = \Gamma(u,u)$], it is clear that each $\varphi_n(u) \propto \Delta^{-1/2}$. By dimensional analysis then, one may derive that $\varphi_{qn} = O(\Delta^{q+1/2})$.

First, we analyze the asymptotic behavior of $\mathcal{N}_0$. According to Eq.~(\ref{n1valexp}), $\mathcal{N}_0$ is a superposition of the products $\varphi_{qn}\varphi_{pn}^* = O(\Delta^{q+p+1})$. Therefore, the $q=p=0$ term in Eq.~(\ref{n1valexp}) dominates in the limit $\Delta \rightarrow 0$ and we find that
\begin{align}
    \mathcal{N}_0 = O(\Delta) \quad \text{and} \quad \mathcal{N}_0^{-1/2} = O(\Delta^{-1/2}). \label{simpleN1}
\end{align}
For an arbitrary $\text{Re}(\varphi_{rm})\in\mathcal{P}$, we find, from Eq.~(\ref{n1valexp}), that
\begin{align}
    \frac{\partial \mathcal{N}_0}{\partial \text{Re}(\varphi_{rm})} &= \sum_{q=0}^\infty \left( \frac{H_{rq}}{r!q!}\varphi_{qm}^* + \frac{H_{qr}}{q!r!}\varphi_{qm} \right)\\
    &= \sum_{q=0}^\infty \frac{2\text{Re}\left( H_{qr} \varphi_{qm} \right)}{r!q!}, \label{dervN1}
\end{align}
where we used the Hermiticity of $H_{qr}$. If one were to have taken $\text{Im}(\varphi_{qn}) \in \mathcal{P}$ instead, the real part in Eq.~(\ref{dervN1}) simply becomes the imaginary part. Equation~(\ref{dervN1}) is a superposition of terms of terms that are $O(\Delta^{q+1/2})$ for all values of $q$. Therefore, the $q=0$ dominates asymptotically and
\begin{align}
    \frac{\partial \mathcal{N}_0}{\partial \varphi_{rm}} &= O(\Delta^{1/2}). \label{n1derv}
\end{align}
Using the chain rule, one finds that
\begin{align}
    \frac{\partial \mathcal{N}_0^{-1/2}}{\partial \varphi_{rm}} = -\frac{\mathcal{N}_0^{-3/2}}{2}\frac{\partial \mathcal{N}_0}{\partial \varphi_{rm}} = O(\Delta^{-1}),
\end{align}
as desired. 

Lastly, we look at the asymptotic behavior of $\partial \varphi /\partial \varphi_{qn}$. Note that $\varphi$ is to be interpreted as a matrix, and its asymptotic behavior references all of its elements. Note however that the elements of $\varphi$ are simply those in $\mathcal{P}$ and therefore
\begin{align}
    \frac{\partial \varphi_{qn}}{\partial \text{Re}(\varphi_{rm})} = \delta_{nm} \delta_{qr} \quad \text{and} \quad \frac{\partial \varphi_{qn}}{\partial \text{Im}(\varphi_{rm})} = \ii \delta_{nm} \delta_{qr} ,
\end{align}
where we use the Kronecker symbol $\delta$. Hence, we arrive at the simple relation that
\begin{align}
    \frac{\partial \varphi}{\partial \varphi_{rm}} = O(1).
\end{align}

\subsection{Asymptotic behavior for the quantum FI bound}
In this section, we show why
\begin{align}
    \tilde{Q}_\alpha (\varphi_{qn}) \sim 4C_\alpha (\varphi_{qn}), \label{Bgoneapp}
\end{align}
for all $\alpha, q,n$. That is, we show why $B_{\alpha}(\varphi_{qn})^2$ is dominated by $C_{\alpha}(\varphi_{qn})$ asymptotically. Defining, for simplicity, $G \triangleq \mathcal{N}_\alpha^{-1/2} \varphi$, we have that
\begin{align}
    B_{\alpha}(\varphi_{qn}) &=  \Tr \left( \Pi   G \partial_i G^\dagger\right) = \Tr\left[(\partial_i G) G^\dagger \Pi \right] \nonumber\\
    &= \Tr \left[\Pi (\partial_i G) G^\dagger \right],
\end{align}
where the second equality comes from taking the transpose of the quantity within the trace, noting that $\Pi^\dagger = \Pi$, which is a result of the assumption that the optical transfer function of the system is even. Then, we note that
\begin{align}
    \Tr(\hat{\rho}_\alpha) = \Tr\left(\Pi G G^\dagger \right),
\end{align}
and therefore, by the product rule
\begin{align}
    \frac{\partial \Tr(\hat{\rho}_\alpha)}{\partial \varphi_{qn}} &= \Tr \left[\Pi \left(\frac{\partial G}{\partial \varphi_{qn}} \right) G^\dagger \right] + \Tr \left(\Pi  G \frac{\partial G^\dagger}{\partial \varphi_{qn}}  \right) \nonumber\\
    &= 2B_{\alpha}(\varphi_{qn}).
\end{align}
We now see that the quantity $B_\alpha$ is related to the rate of change of $\Tr(\hat{\rho}_\alpha)$ w.r.t a parameter. Note that $\Tr(\hat{\rho}_0) = 1$ is a constant, so $B_0(\varphi_{qn}) = 0$. Hence, for IN, Eq.~(\ref{Bgoneapp}) becomes an equality. For ON, note that $\Tr(\hat{\rho}_1) = \mathcal{N}_0$, which is not a constant in general, w.r.t the parameters in $\mathcal{P}$. Therefore,
\begin{align}
    B_1(\varphi_{qn}) &= \frac{1}{2} \frac{\partial \mathcal{N}_0}{\partial \varphi_{qn}} = O(\Delta^{1/2}), \label{Bdep}
\end{align}
where we used Eq.~(\ref{n1derv}). The results in the main body show that $C_1(\varphi_{qn}) = O(1)$, which dominates $B_1(\varphi_{qn})^2 = O(\Delta)$ asymptotically. 

In summary, $B_\alpha$ is the rate of change of the trace of the density matrix. For IN ($\alpha = 0$), the trace is constant and therefore $B_0$ vanishes, leading to Eq.~(\ref{Bgoneapp}). For ON ($\alpha = 1$), the trace varies w.r.t to the parameters in $\mathcal{P}$ and so $B_1 \neq 0$. Nevertheless, in the sub-Rayleigh regime, it turns out that $C_1$ dominates $B_1$, which once again leads to Eq.~(\ref{Bgoneapp}).

\section{Parametrization with intensity moments} \label{appF}
In this section, we derive a relation between $\theta_q$ and $\varphi_{qn}$ so that the reparametrization property of the quantum FIB matrix may be used to obtain quantum bounds for $\theta_q$. For convenience, we define
\begin{align}
    \Phi_{qn} \triangleq \int_{-\infty}^\infty |\varphi_n (u)|^2  u^q \, \ud u. \label{bigPhi}
\end{align}
With this, the CMD gives
\begin{align}
    \theta_q = \sum_{n=0}^\infty \Phi_{qn}. \label{CMDtheta}
\end{align}
We now restrict ourselves to the case where $q$ is even. Here, we find that Eq.~(\ref{bigPhi}) may be written in a form on which the Plancherel theorem may be applied:
\begin{align}
    \Phi_{qn} &= \int_{-\infty}^\infty |\varphi_n(u) u^{q/2}|^2 \, \ud u \nonumber\\
    &= \frac{1}{( 2\pi)^{q}} \int_{-\infty}^\infty \left| \frac{\partial^{(q/2)} \tilde{\varphi}_n (f)}{\partial f^{(q/2)} } \right|^2 \, \ud f, \label{bigPhi2}
\end{align}
where $\tilde{\varphi}_n(f)$ is defined as the Fourier transform of $\varphi_n(u)$, which is assumed to be smooth:
\begin{align}
    \tilde{\varphi}_n(f) &= \int_{-\infty}^\infty \varphi_n(u) e^{\ii 2\pi f u}\, \ud u \nonumber\\
    &= \sum_{k=0}^\infty \frac{(-\ii 2\pi f)^k}{k!} \int_{-\infty}^\infty \varphi_n(u) u^k \, \ud u \nonumber\\
    &= \sum_{k=0}^\infty \frac{(-\ii 2\pi f)^k}{k!} \varphi_{kn}. \label{fourierPhi}
\end{align}
Note that the existence of $\tilde{\varphi}_n(f)$ ensures the convergence of Eq.~(\ref{fourierPhi}) for all $f$. Therefore, we find that
\begin{align}
    \frac{\partial^{(q/2)} \tilde{\varphi}_n(f)}{\partial f^{(q/2)} } = \sum_{k=q/2}^\infty \frac{(-\ii 2\pi f)^{k-q/2}}{(k-q/2)!} \varphi_{kn}. \label{diff}
\end{align}
Equation~(\ref{diff}) may now be inserted into Eq.~(\ref{bigPhi2}) to obtain
\begin{align}
    \Phi_{qn} &= \frac{1}{(2\pi)^q}\int_{-\infty}^\infty \left[ \sum_{k_1,k_2 = q/2}^\infty H_q^{k_1 k_2}(f) \varphi_{k_1 n} \varphi_{k_2 n}^* \right] \, \ud f 
\end{align}
with
\begin{align}
    H_q^{k_1 k_2} \triangleq  (-1)^{k_1 - q/2}\frac{(\ii 2 \pi f)^{k_1 + k_2 - q}}{(k_1-q/2)! (k_2 - q/2)!}. \label{infinity}
\end{align}
Note that $H^{k_1,k_2}_q = (H^{k_2 k_1}_q)^*$. Returning to Eq.~(\ref{CMDtheta}), we find that
\begin{align}
    \theta_q &= \frac{1}{(2\pi)^q}\sum_{n=0}^\infty  \int_{-\infty}^\infty \left[ \sum_{k_1,k_2 = q/2}^\infty \!\!\!\!H_q^{k_1 k_2}(f) \varphi_{k_1 n} \varphi_{k_2 n}^* \right] \! \ud f  , \label{varphitothetaapp}
\end{align}
which, once again, is valid for even $q$. Although one may be concerned with the convergence of Eq.~(\ref{varphitothetaapp}) since Eq.~(\ref{infinity}) appears to diverge (upon integration) for any $k_1,k_2$, we note that Eq.~(\ref{varphitothetaapp}) includes a summation over $k_1,k_2$, which is to be done prior to the integral.

For odd $q$, we note that Eq.~(\ref{bigPhi2}) is invalid since $|u|^q \neq u^q$ and a new interpretation of derivatives is needed. Instead, we write
\begin{align}
    \Phi_{qn} \!&= \! \int_\infty^\infty\!\! |\varphi_n(u) u^{q/2} \Theta(u)|^2 \, \ud u \nonumber\\
    &-\int_{-\infty}^\infty\!\! |\varphi_n(u)u^{q/2}\Theta(-u)|^2 \, \ud u, \label{oddq}
\end{align}
where $\Theta(u)$ is the Heaviside step function defined as
\begin{align}
    \Theta(u) = \begin{cases} 1 & u\ge0, \\ 0 &u <0. \end{cases}
\end{align}
Let us define $\tilde{\Theta}(f)$ to be the Fourier transform of $\Theta(u)$ and further take $\mathcal{H}(f) \triangleq \tilde{\Theta}(f) - \tilde{\Theta}(-f)$. We now show that the derivations for even $q$ have an analogous result for odd $q$. With Eq.~(\ref{oddq}), the quantity within the mod-square in Eq.~(\ref{bigPhi2}) becomes: 
\begin{align}
     \frac{\partial^{(q/2)} \tilde{\varphi}_n (f)}{\partial f^{(q/2)} } * \mathcal{H}(f), \label{oddqconv}
\end{align}
where $*$ denotes a convolution. Recalling that $q/2$ is not an integer, one must re-interpret the derivatives in Eq.~(\ref{oddqconv}) properly. For the purposes of consistency with the Fourier transform property of derivatives, one may consider the Caputo fractional derivative:
\begin{align}
    \frac{\partial^{(q/2)} \tilde{\varphi}_n(f)}{\partial f^{(q/2)}} \triangleq \frac{\partial^{\lceil q/2 \rceil} \tilde{\varphi}_n(f)}{\partial f^{\lceil q/2 \rceil}} * \left[\frac{1}{\Gamma(1/2)} f^{-1/2} \right], \label{caputo}
\end{align}
where $\Gamma$ is the gamma function (and not the mutual coherence, which commonly shares the same symbol). With Eq.~(\ref{caputo}), one may follow similar derivations for even $q$ above to show that
\begin{align}
    \theta_q &= \frac{1}{(2\pi)^q}\sum_{n=0}^\infty  \int_{-\infty}^\infty \left[ \sum_{k_1,k_2 = \lceil q/2 \rceil}^\infty \!\!\!\! A_q^{k_1 k_2}(f) \varphi_{k_1 n} \varphi_{k_2 n}^* \right]  \ud f  ,
\end{align}
which is valid for all $q$. Note that, suppressing explicit $f$-dependence,
\begin{align}
    A_q^{k_1 k_2} \!&=\! \begin{cases}
     H_q^{k_1 k_2}  & q \text{ even},\\
    H_q^{k_1 k_2}* \left[ \Gamma^{-1}\left(\frac{1}{2} \right) f^{-1/2}\right] * \mathcal{H} & q \text{ odd},
    \end{cases}
\end{align}
and $A^{k_1,k_2}_q = (A^{k_2 k_1}_q)^*$.

\nocite{*}

\bibliography{Main}

\end{document}